\documentclass{doublecol-new}

\usepackage{natbib,stfloats}
\usepackage{mathrsfs}
\usepackage{graphicx}
\usepackage{textcomp}
\usepackage{makecell}
\usepackage{pgfplots}
\usepackage{amsmath}
\usepackage{tabto}
\usepackage{wrapfig}
\usepackage{amssymb}
\usepackage{natbib}
\usepackage{float}

\usepackage{caption}
\usepackage{xcolor,listings}
\theoremstyle{TH}{

}

\theoremstyle{THrm}{

}

\theoremstyle{THhit}{

}

\makeatletter
\def\theequation{\arabic{equation}}

\JOURNALNAME{\TEN{\it Int. J. Business Process
Integration and Management}}%

\makeatother

\begin{document}%

\setcounter{page}{1}

\LRH{Ngo, V.M., Le-Khac, N.A. and Kechadi M.T.}
\RRH{Int. J. Business Process
Integration and Management}
\VOL{10}
\ISSUE{1}
\PUBYEAR{2020}


\title{Data Warehouse and Decision Support on \\ Integrated Crop Big Data}

\authorA{Vuong M. Ngo}
\affA{E-mail: vuong.nm@ou.edu.vn, vuong.ngo@ucd.ie}

\authorB{Nhien-An Le-Khac}
\affB{E-mail: an.lekhac@ucd.ie}

\authorC{M-Tahar Kechadi}
\affC{E-mail: tahar.kechadi@ucd.ie
\newline
\\Ho Chi Minh City Open University, HCMC, Vietnam
\\University College Dublin, Belfield, Dublin 4, Ireland}

\begin{abstract}
  In recent  years, precision agriculture  is becoming  very popular. The  introduction of
  modern  information  and  communication   technologies  for  collecting  and  processing
  Agricultural data revolutionise the agriculture practises.  This has started a while ago
  (early  20th century)  and  it  is driven  by  the low  cost  of  collecting data  about
  everything; from information on fields such  as seed, soil, fertiliser, pest, to weather
  data, drones  and satellites images.  Specially, the  agricultural data mining  today is
  considered as Big  Data application in terms of volume,  variety, velocity and veracity.
  Hence  it  leads  to challenges  in  processing  vast  amounts  of complex  and  diverse
  information  to  extract  useful  knowledge   for  the  farmer,  agronomist,  and  other
  businesses. It is  a key foundation to establishing a  crop intelligence platform, which
  will enable efficient resource management and  high quality agronomy decision making and
  recommendations.  In  this  paper,  we  designed and  implemented  a  continental  level
  agricultural data warehouse (ADW). ADW is  characterised by its (1) flexible schema; (2)
  data integration  from real agricultural multi  datasets; (3) data science  and business
  intelligent  support;  (4)  high  performance;  (5)  high  storage;  (6)  security;  (7)
  governance and  monitoring; (8)  consistency, availability  and partition  tolerant; (9)
  cloud compatibility.  We also evaluate the  performance of ADW and  present some complex
  queries to extract and return necessary knowledge about crop management.
\end{abstract}

\KEYWORD{Data warehouse, decision support, crop Big Data, smart agriculture.}

\REF{to this paper should  be made as follows: Ngo, V.M., Le-Khac,  N.A. and Kechadi, M.T.
  (2020)  `Data  Warehouse  and  Decision  Support on  Integrated  Crop  Big  Data',  {\it Int. J. Business Process Integration and Management}, Vol. 10, No. 1,
  pp. 17\textendash 28.} 

\begin{bio}
Vuong M. Ngo received the B.E, M.E and  PhD degrees in computer science at HCMC University
of Technology in 2004, 2007 and 2013  respectively. He is currently a Senior Researcher at UCD and HCMC Open University. His  research interests
include information  retrieval, sentiment analysis,  data mining, graph matching  and data
\newline
\newline
\noindent Nhien-An Le-Khac is currently a Lecturer  at the School of Computer Science, UCD
and  a  Programme  Director  of  MSc   programme  in  forensic  computing  and  cybercrime
investigation. He obtained  his PhD in computer  science in 2006 at  the Institut National
Polytechnique Grenoble, France. His research interest  spans the area of cybersecurity and
digital  forensics,  data mining/distributed  data  mining  for  security, grid  and  high
performance computing. 
\newline
\newline
\noindent M-Tahar  Kechadi was  awarded PhD  and Master degrees  in computer  science from
University of Lille 1, France. He joined the UCD School of Computer Science in 1999. He is
currently Professor of Computer  Science at UCD. His research interests  span the areas of
data mining, data  analytics, distributed data mining,  heterogeneous distributed systems,
grid  and  cloud Computing,  cybersecurity,  and  digital  forensics.  He is  a  Principal
Investigator at Insight  Centre for Data Analytics  and CONSUS project. He is  a member of
IEEE and ACM.
 \end{bio}

\maketitle

\section{Introduction}
\label{sec:Intro}
Annual world  cereal productions  were $2,608$  million tons and  $2,595$ million  tons in
$2017$ and $2018$, respectively \citep{USDA.2018, FAO-CSDB.2018}. However, there were also
around  $124$ million  people in  $51$  countries faced  food crisis  and food  insecurity
\citep{FAO-FSIN.2018}. According to United  Nations \citep{UnitedNations.2017}, we need an
increase $60\%$  of cereal  production to meet  $9.8$ billion people  needs by  $2050$. To
satisfy the  huge increase demand  for food, crop  yields must be  significantly increased
using modern farming approaches, such as  smart farming also called precision agriculture.
As  highlighted in  the European  Commission report  \citep{Eurobarometer.2016}, precision
agriculture is vitally important for the future and can make a significant contribution to
food security and safety.

The precision  agriculture's current mission is  to use the decision-support  system (DSS)
based on Big Data approaches to provide  precise information for more control of waste and
farming   efficiency,  such   as  soil   nutrient  \citep{Rogovska.2019},   early  warning
\citep{Rembold.2019},     forecasting      \citep{Bendre.2015},     irrigation     systems
\citep{Huang.2013},   evapotranspiration    prediction   \citep{Paredes.2014}, soil and herbicide, insecticide optimisation \citep{Ngo:2020}, awareness
\citep{Lokers.2016},   supply   chain   \citep{Protopop.2016} and financial   services
\citep{Ruan.2019}. Normally, the DSSs implement  a knowledge discovery process also called
data  mining  process,  which  consists  of  data  collection  and  data  modelling,  data
warehousing,  data  analysis  (using  machine learning  or  statistical  techniques),  and
knowledge deployment  \citep{Dicks.2014}. Hence,  designing and implementing  an efficient
agricultural data warehouse (ADW) is one of the key steps of this process, as it defines a
uniform data  representation through its schema  model and stores the  derived datasets so
that they can be  analysed to extract useful knowledge. However,  currently, this step was
not given  much attention. Therefore,  there are very few  reports in the  literature that
focus  on the  design of  efficient ADWs  with the  view to  enable Agricultural  Big Data
analytics and  mining. The design  of large scale ADWs  is very challenging.  Because, the
agricultural  data is  spatial, temporal,  complex, heterogeneous,  non-standardised, high
dimensional, collected from  multi-sources, and very large. In particular,  it has all the
features of  Big Data;  volume, variety,  velocity and  veracity. Moreover,  the precision
agriculture system can be used by different kinds  of users at the same time, for instance
by farmers,  policymakers, agronomists,  and so on.  Every type of  user needs  to analyse
different information, sets thus requiring specific analytics.

Unlike  in  any  other  domains;  health-care,  financial data,  etc,  the  data  and  its
warehousing in precision  agriculture are unique. This is because,  there are very complex
relationships  between  the  agricultural  data  dimensions. The  data  sources  are  very
diversified and varying levels of quality. Precision agriculture (PA) warehousing has many
decision-making processes  and each needs  different levels  of data access  and different
needs of analysis. Finally, there are many stakeholders involved in the data ownership and
exploitation. So,  the data  has significant  number of  uncertainties. For  examples, the
quality of  data collected by  farmers depends directly  on their knowledge,  routines and
frequency of information recording,  and support tools, etc. All these  issues make the PA
data unique when it  becomes to its storage, access, and analysis.  These issues may exist
in other domains, but not at the same scale and as in agriculture practices.

In  this research,  we firstly analyse  real-world agricultural  Big Data  to build  the
effective constellation  schema. From  this schema,  some simple  questions can  be easily
answered directly from the modelled data. These  questions include: (1) For a given field,
what kind of crops are suitable to grow?  (2) Which companies can purchase a specific crop
with the  highest price  in the  past season?  (3) List  the history  of soil  texture and
applied fertilisers for a  given field; (4) List costs of production  for wheat and barley
in the last  5 years, and so on.  Secondly, the proposed ADW has enough  main features and
characteristics of  Big Data Warehouse  (BDW). These are  (1) high storage  capacity, high
performance and cloud computing compatibility;  (2) flexible schema and integrated storage
structure; (3)  data ingestion, monitoring, and  security to deal with  the data veracity.
Besides, an experimental evaluation is conducted to study the performance of ADW storage. 

The rest  of this  paper is organised  as follows:  in the next  Section, we  reviewed the
related  work about  decision  support  systems and  data  warehouses  in agriculture.  In
Sections \ref{sec:CBD}, \ref{sec:DW} and \ref{sec:EO}, we presented big data aspects of PA, our ADW architecture and its modules. In Sections \ref{sec:QC}, \ref{sec:ADWI}, \ref{sec:PA} and \ref{sec:ADM}, the quality criteria, implementation, performance analysis and decision-making applications of the proposed ADW are presented respectively. Section \ref{sec:CFW} gives some concluding remarks and  future research  directions.  Finally,  a concrete  example  about the ADW and its operational average run-times are shown in the appendix.

\section{Related Work}
\label{sec:RW}
In precision  agriculture, DSSs  are designed  to support  different stakeholders  such as
farmers, advisers  and policymakers to  optimise resources, farms' management  and improve
business practices  \citep{Gutierreza.2019}. For  instance, DSSs were  built to  1) manage
microbial  pollution  risks in  dairy  farming  \citep{Oliver.2017}; 2)  analyse  nitrogen
fertilisation from  satellite images \citep{Lund_Lind.2018};  3) control pest  and disease
under uncertainty in climate conditions \citep{Devitt.2017}; 4) manage drip irrigation and
its    schedule   \citep{Friedman.2016};    5)   predict    and   adopt    climate   risks
\citep{Han.2017}.  However, the  datasets  that were  used in  the  mentioned studies  are
small.  Besides,  they focused  on  using  visualisation  techniques to  assist  end-users
understand and interpret their data.

Recently, many  papers have  been published  on how to  exploit intelligent  algorithms on
sensor  data  to  improve agricultural  economics  \cite{Pantazi.2016},  \cite{Park.2016},
\cite{Hafezalkotob.2018},      \cite{Udiasa.2018}     and      \cite{Rupnik.2019}.      In
\cite{Pantazi.2016},  the  authors predicted  crop  yield  by using  self-organising-maps;
namely   supervised  Kohonen   networks,  counter-propagation   artificial  networks   and
XY-fusion. In \cite{Park.2016}, one predicted drought conditions by using three rule-based
machine learning; namely  random forest, boosted regression trees, and  Cubist.  To select
the best  olive harvesting  machine, the authors  in \cite{Hafezalkotob.2018}  applied the
target-based  techniques on  the main  criteria,  which are  cost, vibration,  efficiency,
suitability,  damage,  automation, work  capacity,  ergonomics,  and safety.   To  provide
optimal  management of  nutrients and  water, the  paper \cite{Udiasa.2018}  exploited the
multi-objective genetic  algorithm to implement  an E-Water system.  This  system enhanced
food crop  production at  river basin  level. Finally,  in \cite{Rupnik.2019}  the authors
predicted pest population  dynamics by using time series clustering  and structural change
detection  which  detected  groups  of  different pest  species.   However,  the  proposed
solutions are not scalable enough to handle agricultural Big Data; they present weaknesses
in one of the following aspects: data integration, data schema, storage capacity, security
and performance.

From a  Big Data point of  view, the papers \cite{Kamilaris.2018}  and \cite{Schnase.2017}
have proposed “smart agricultural frameworks”.  In \cite{Kamilaris.2018}, the authors used
Hive to store  and analyse sensor data  about land, water and biodiversity  which can help
increase  food production  with  less environmental  impact.  In \cite{Schnase.2017},  the
authors  moved  toward   a  notion  of  climate  analytics-as-a-service,   by  building  a
high-performance analytics and scalable data management platform, which is based on modern
cloud infrastructures, such as Amazon web services, Hadoop, and Cloudera. However, the two
papers did not discuss how to build and implement a DW for a precision agriculture.

The   proposed   approach,   inspired   from   \cite{Schulze.2007},   \cite{Schuetz.2018},
\cite{Nilakanta.2008} and  \cite{Ngo.2018}, introduces ways of  building agricultural data
warehouse (ADW). In \cite{Schulze.2007},  the authors extended entity-relationship concept
to  model operational  and analytical  data; called  multi-dimensional entity-relationship
model.  They also introduced new representation elements and showed how can be extended to
an analytical  schema. In  \cite{Schuetz.2018}, a  relational database  and an  RDF triple
store were proposed to  model the overall datasets. The data is loaded  into the DW in RDF
format,  and cached  in the  RDF  triple store  before being  transformed into  relational
format.   The  actual   data  used   for  analysis   was  contained   in  the   relational
database. However, as the schemas used in \cite{Schulze.2007} and \cite{Schuetz.2018} were
based on entity-relationship models, they cannot  deal with high-performance, which is the
key feature of a data warehouse.

In \cite{Nilakanta.2008}, a star schema model was used. All data marts created by the star
schemas are  connected via  some common dimension  tables. However, a  star schema  is not
enough to present complex agricultural information and  it is difficult to create new data
marts   for  data   analytics.  The   number  of   dimensions  of   the  DW   proposed  in
\cite{Nilakanta.2008}  is  very   small;  only  3-dimensions  –   Species,  Location,  and
Time. Moreover, the  DW concerns livestock farming.  Overcoming disadvantages  of the star
schema, the authors of \cite{Ngo.2018} and \cite{Ngo:2020} proposed a constellation schema for an agricultural
DW architecture in order  to satisfy the quality criteria. However,  they did not describe how to design and implement their DW.

\vspace{-1mm}
\section{Crop Big Data}
\label{sec:CBD}
\vspace{-1mm}
\subsection{Crop Datasets}

\begin{figure}[H]
  \begin{center}
    \includegraphics[width=0.36\textwidth]{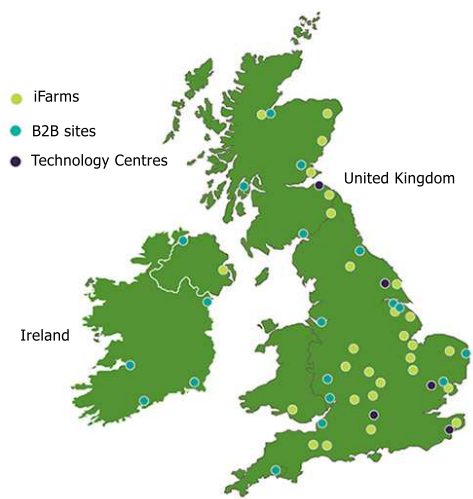}
 	\caption{Data from UK and Ireland. 
 	}
 	\vspace{-2mm}
	\label{fig_UKIreland}
  \end{center}
\end{figure}

\begin{figure}[H]
  \vspace{0 mm}
  \includegraphics[width=0.43\textwidth]{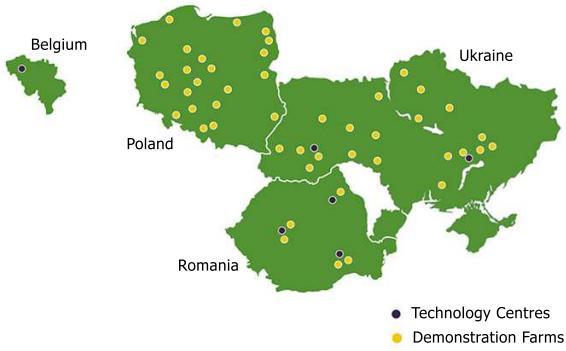}
  \caption{Data in Continental Europe. 
  }
  \label{fig_continentaleurope}
  \vspace{-2 mm}
\end{figure}

The datasets  were primarily obtained  from an agronomy  company, which extracted  it from
them operational data storage systems, research  results, and field trials. Especially, we
were given real-world  agricultural datasets on iFarms,  Business-to-Business (B2B) sites,
technology centres  and demonstration farms.  Theses datasets were collected  from several
European   countries  and   they  are   presented  in   Figures  \ref{fig_UKIreland}   and
\ref{fig_continentaleurope} \citep{Origin.2018}.  These datasets  describe more than $112$
distribution points, $73$ demonstration farms, $32$ formulation and processing facilities,
$12.7$ million hectares of direct farm customer footprint and $60,000$ trial units. 

\vspace{-6 mm}
\vspace{-10mm}

There is  a total of  29 datasets. On  average, each dataset  contains $18$ tables  and is
about $1.4$ GB in  size.  Each dataset focuses on a few information  that impact the crop.
For instance,  the weather dataset includes  information on location of  weather stations,
temperature, rainfall and wind speed over  time.  Meanwhile, soil component information in
farm sites, such  as mineral, organic matter, air, water  and micro-organisms, were stored
in the  soil dataset.  The  fertiliser dataset contains  information about field  area and
geographic position, crop name, crop yield, season, fertiliser name and quantity.

\subsection{Big Data Challenges}
Raw  and  semi-processed  agricultural  datasets are  usually  collected  through  various
sources: Internet of  Thing (IoT) devices, sensors, satellites,  weather stations, robots,
farm  equipment, farmers  and agronomists,  etc. Besides,  agricultural datasets  are very
large, complex, unstructured, heterogeneous, non-standardised, and inconsistent. Hence, it
has all the features of Big Data.
\begin{enumerate}
  \item {\bf Volume:}  The  amount of  agricultural  data is  increasing  rapidly and  is
        intensively produced by  endogenous and exogenous sources. The  endogenous data is
        collected  from  operational  systems,   experimental  results,  sensors,  weather
        stations,  satellites, and  farming  equipment.  The systems  and  devices in  the
        agricultural ecosystem can  be connected through IoT. The  exogenous data concerns
        the external  sources, such as  government agencies, retail agronomists,  and seed
        companies.  They can  help with information about local pest  and disease outbreak
        tracking, crop monitoring, food security, products, prices, and knowledge.
  \item {\bf Variety:} Agricultural data has  many different forms and formats, structured
        and unstructured  data, video, imagery, chart,  metrics, geo-spatial, multi-media,
        model, equation, text, etc.
  \item {\bf  Velocity:} The collected  data increases at very  high rate, as  sensing and
        mobile  devices are  becoming more  efficient and  cheaper. The  datasets must  be
        cleaned, aggregated and harmonised in real-time.
  \item {\bf  Veracity:} The  tendency  of  agronomic  data is  uncertain,  inconsistent,
        ambiguous and error prone because the data is gathered from heterogeneous sources,
        sensors and manual processes.
\end{enumerate}


\subsection{ADW Schema}

\begin{figure}[H]
  \begin{center}
    \includegraphics[width=0.48\textwidth, height=0.76\textheight]{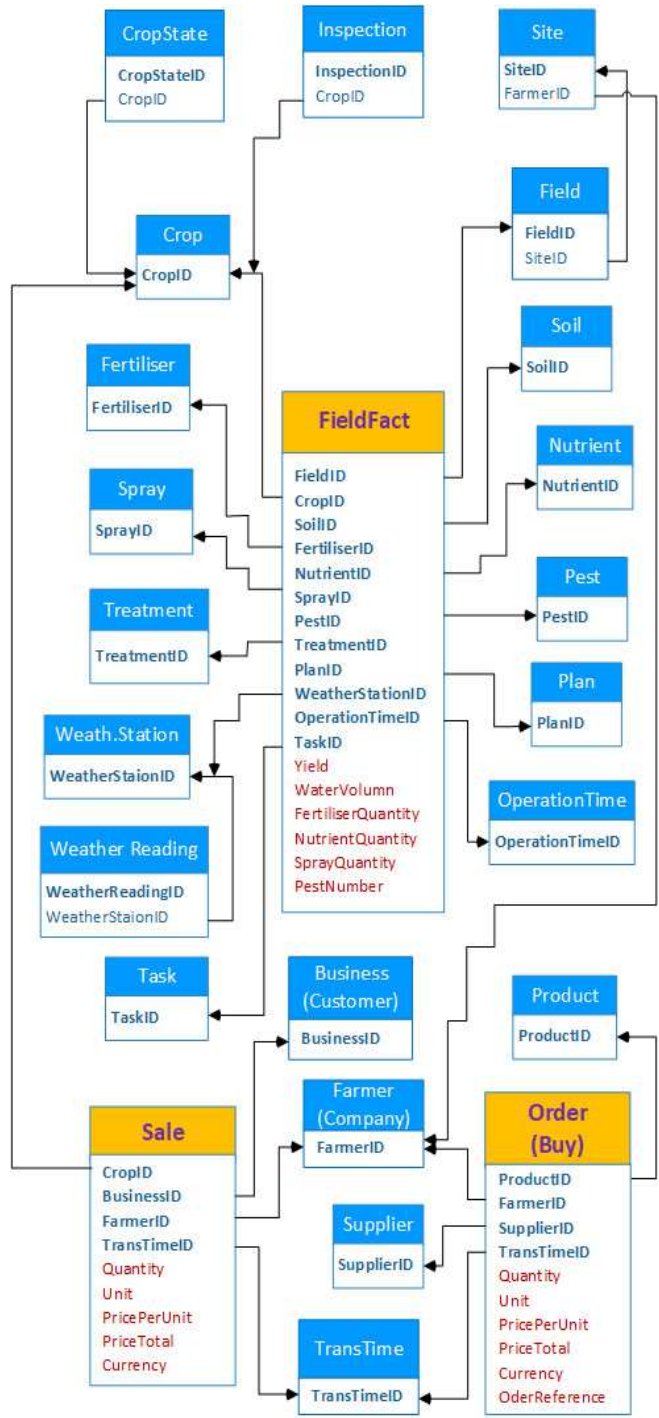}
  \end{center}
  \vspace{-2mm}
  \caption{A part of ADW schema for Precision Agriculture}
  \label{fig_Schema}
\end{figure}	

The DW uses schema to logically describe the  entire datasets. A schema is a collection of
objects, including  tables, views, indexes,  and synonyms which  consist of some  fact and
dimension tables \citep{Oracle.2017}. The DW schema can  be designed based on the model of
source data  and the  user requirements.   There are  three kind  of models,  namely star,
snowflake and fact constellation. With the its  various uses, the ADW schema needs to have
more than one fact table and should  be flexible. So, the constellation schema, also known
galaxy schema should be used to design the ADW schema.

\begin{figure}[H]
  \begin{center}
    \includegraphics[width=0.33\textwidth, height=0.25\textheight]{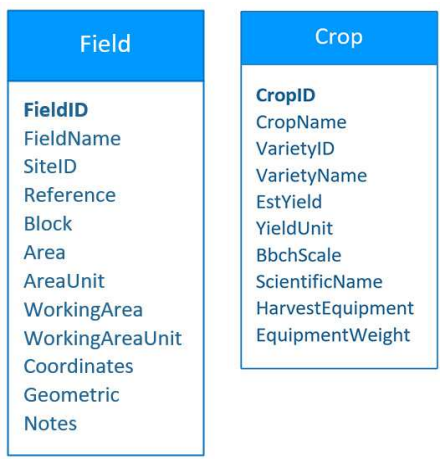}
  \end{center}
  \vspace{-4mm}
  \caption{Field and Crop dimension tables}
  \vspace{-5mm}
  \label{fig_dimension1}
\end{figure}

\begin{figure}[H]
  \begin{center}
    \includegraphics[width=0.36\textwidth, height=0.29\textheight]{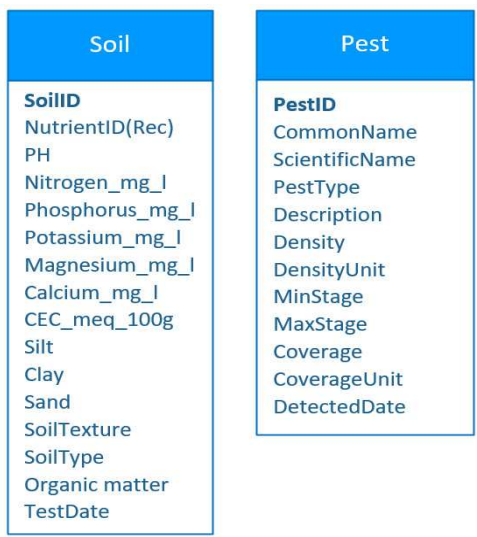}
  \end{center}
  \vspace{-4mm}
  \caption{Soil and Pest dimension tables}
  \vspace{-3mm}
  \label{fig_dimension2}
\end{figure}

We  developed a  constellation schema  for ADW  and it  is partially  described in  Figure
\ref{fig_Schema}.  It includes few fact tables  and many dimension tables. {\tt FieldFact}
fact table  contains data about  agricultural operations on  fields. {\tt Order}  and {\tt
  Sale} fact  tables contain  data about  farmers' trading  operations. The  key dimension
tables are  connected to their  fact table. There are  some dimension tables  connected to
more than one  fact table, such as  {\em Crop} and {\em Farmer}.  Besides, {\em CropState,
  Inspection, Site}, and  {\em Weather Reading} dimension tables are  not connected to any
fact table.  {\em CropState} and  {\em Inspection} tables are  used to support  {\em Crop}
table.  While, {\em  Site} and {\em Weather  Reading} tables support {\em  Field} and {\em
  WeatherStation} tables. {\tt FieldFact} fact table  saves the most important facts about
teh field; yield, water volume, fertiliser quantity, nutrient quantity, spray quantity and
pest number.  While, in  {\tt Order} and {\tt Sale} tables, the  important facts needed by
farm management are quantity and price.

\begin{table}[H]
 \begin{center}
  \vspace{-4mm}
  \caption{Descriptions of other dimension tables}
  \vspace{0mm}
    \small
    \begin{tabular}{|m{4mm}|m{14mm}|m{52mm}|}
      \hline
      \textbf{No.} & \textbf {Dim. tables} & \textbf{Particular attributes}\\
      \hline
      1 & Business & BusinessID, Name, Address, Phone, Mobile, Email\\
      \hline
      2 & CropState & CropStateID, CropID, StageScale, Height, MajorStage, MinStage, MaxStage, Diameter, MinHeight, MaxHeight, CropCoveragePercent \\
      \hline
      3 & Farmer & FarmerID, Name, Address, Phone, Mobile, Email \\
      \hline
      4 & Fertiliser & FertiliserID, Name, Unit, Status, Description, GroupName \\	
      \hline
      5 & Inspection & InspectionID, CropID, Description, ProblemType, Severity, Problem\-Notes, AreaValue, AreaUnit, Order, Date, Notes, GrowthStage \\
      \hline
      6 & Nutrient & NutrientID, NutrientName, Date, Quantity \\
      \hline
      7 & Operation Time & OperationTimeID, StartDate, End\-Date, Season \\
      \hline
      8 & Plan & PlanID, PName, RegisNo, Product\-Name, ProductRate, Date, Water\-Volume \\
      \hline
      9 & Product & ProductID, ProductName, Group\-Name \\
      \hline
      10 & Site & SiteID, FarmerID, SiteName, Reference, Country, Address, GPS, CreatedBy\\
      \hline
      11 & Spray & SprayID, SprayProductName, ProductRate, Area,Date, WaterVol, ConfDuration, ConfWindSPeed, ConfDirection, ConfHumidity, Conf\-Temp, ActivityType \\
      \hline
      12 & Supplier & SupplierID, Name, ContactName, Address, Phone, Mobile, Email\\
      \hline
      13 & Task & TaskID, Desc, Status, TaskDate, TaskInterval, CompDate, AppCode \\
      \hline
      14 & Trans Time & TransTimeID, OrderDate, Deliver\-Date, ReceivedDate, Season \\
      \hline
      15 & Treatment & TreatmentID, TreatmentName, FormType, LotCode, Rate, Appl\-Code, LevlNo, Type, Description, ApplDesc, TreatmentComment \\
      \hline
      16 & Weather Reading & WeatherReadingID, WeatherSta\-tionID, ReadingDate, Reading\-Time, AirTemperature, Rainfall, SPLite, RelativeHumidity,  WindSpeed, WindDirection, SoilTemperature, LeafWetness \\
      \hline
      17 & Weather Station & WeatherStationID,
                             StationName, Latitude, Longitude, Region\\
      \hline
    \end{tabular}
    \label{tab5}
  \end{center}
  \vspace{-2mm}
\end{table}

The dimension  tables contain details  on each  instance of an  object involved in  a crop
yield or farm management.  Figure \ref{fig_dimension1} describes attributes of {\em Field}
and {\em Crop} dimension tables. {\em  Field} table contains information about name, area,
co-ordinates (being  longitude and latitude of  the centre point of  the field), geometric
(being a collection of  points to show the shape of the field)  and site identify the site
that the  field it belongs  to. While, {\em Crop}  table contains information  about name,
estimated  yield of  the crop  (estYield), BBCH  Growth Stage  Index (BbchScale),  harvest
equipment and its weight. These provide useful information for crop harvesting.

Figure \ref{fig_dimension2}  describes attributes of  {\em Soil} and {\em  Pest} dimension
tables. {\em Soil} table contains information about PH value (a measure of the acidity and
alkalinity),  minerals  (nitrogen,  phosphorus,  potassium, magnesium  and  calcium),  its
texture (texture  label and percentage of  Silt, Clay and Sand),  cation exchange capacity
(CEC) and  organic matter.   Besides, information about  recommended nutrient  and testing
dates ware also included in this table.  In {\em Pest} table contains name, type, density,
coverage  and detected  dates of  pests. For  the remaining  dimension tables,  their main
attributes are described in Table \ref{tab5}.

 
\section{ADW Architecture}
\label{sec:DW}

A DW  is a federated repository  for all the data  that an enterprise can  collect through
multiple   heterogeneous   data   sources;   internal  or   external.   The   authors   in
\cite{Golfarelli-Rizzi.2009} and \cite{Inmon.2005} defined DW  as a collection of methods,
techniques,  and  tools  used  to  conduct  data  analyses,  make  decisions  and  improve
information resources.  DW is defined around key subjects and involves data cleaning, data
integration and data consolidations. Besides, it must  show its evolution over time and is
not volatile.

\begin{figure*}
  \begin{center}
    \includegraphics[width=0.95\textwidth]{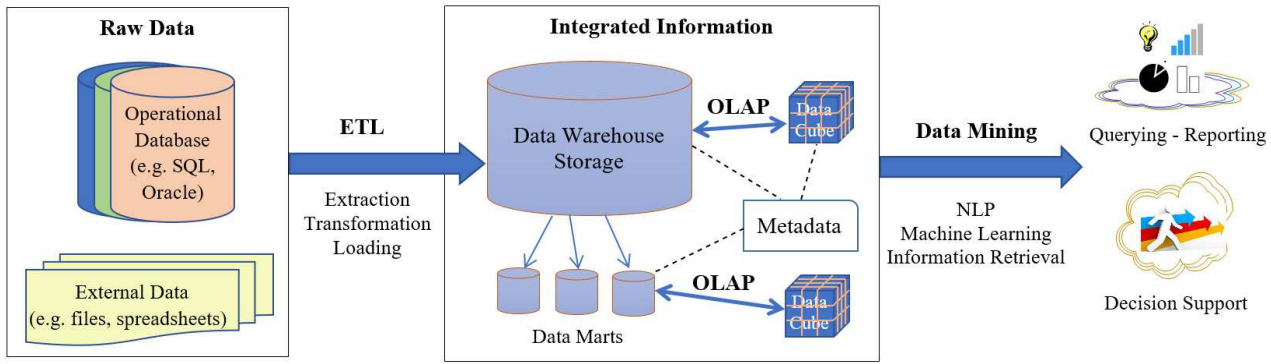}
  \end{center}
  \vspace{-2mm}
  \caption{Agricultural Data Warehouse Architecture.}
  \label{fig_generalDW}
\end{figure*}

The  general architecture  of a  typical  DW system  includes four  separate and  distinct
modules; Raw  Data, Extraction  Transformation Loading  (ETL), Integrated  Information and
Data    Mining    \citep{Kimball-Ross.2013},    which    is    illustrated    in    Figure
\ref{fig_generalDW}.  In that,  Raw  Data (source  data) module  is  originally stored  in
various storage  systems (e.g. SQL, sheets,  flat files, ...). The raw data often requires
cleansing, correcting noise  and outliers, dealing with missing values.   Then it needs to
be integrated and consolidated before loading it into a DW storage through ETL module.

The Integrated  Information module is  a logically centralised repository,  which includes
the DW  storage, data marts,  data cubes  and OLAP engine.   The DW storage  is organised,
stored and  accessed using a  suitable schema  defined by the  metadata. It can  be either
directly accessed or used to create data  marts, which is usually oriented to a particular
business  function or  an  enterprise  department. A  data  mart  partially replicates  DW
storage's contents and is a subset of DW storage. Besides, the data is extracted in a form
of data  cube before  it is analysed  in the  data mining  module. A data  cube is  a data
structure that  allows advanced  analysis of  data according  to multiple  dimensions that
define a given problem. The data cubes are manipulated by the OLAP engine. The DW storage,
data mart and data cube are considered as  metadata, which can be applied to the data used
to define other  data. Finally, Data Mining  module contains a set of  techniques, such as
machine  learning, heuristic,  and statistical  methods  for data  analysis and  knowledge
extraction at multiple level of abstraction.

\section{ETL and OLAP}
\label{sec:EO}

The  ETL module  contains Extraction,  Transformation, and  Loading tools  that can  merge
heterogeneous schemata, extract, cleanse, validate, filter, transform and prepare the data
to be loaded  into a DW. The extraction  operation allows to read, retrieve  raw data from
multiple and different types of data sources  systems and store it in a temporary staging.
During  this operation,  the  data goes  through  multiple checks  --  detect and  correct
corrupted and/or  inaccurate records, such  as duplicate data, missing  data, inconsistent
values and wrong values. The transformation operation structures, converts or enriches the
extracted data and presents  it in a specific DW format. The  loading operation writes the
transformed data  into the DW  storage. The ETL  implementation is complex,  and consuming
significant amount of time and resources. Most DW projects usually use existing ETL tools,
which are classified into two groups.  The  first is a commercial and well-known group and
includes tools  such as  Oracle Data  Integrator, SAP Data  Integrator and  IBM InfoSphere
DataStage. The second  group is famous for  it open source tools, such  as Talend, Pentaho
and Apatar.

OLAP is a category  of software technology that provides the  insight and understanding of
data in multiple  dimensions through fast, consistent, interactive  access, management and
analysis of the  data. By using roll-up (consolidation), drill-down,  slice-dice and pivot
(rotation)  operations, OLAP  performs  multidimensional  analysis in  a  wide variety  of
possible  views of  information that  provides  complex calculations,  trend analysis  and
sophisticated data modelling quickly. The OLAP  systems are divided into three categories:
1)  Relational  OLAP  (ROLAP),  which  uses  relational  or  extended-relational  database
management  system to  store  and  manage the  data  warehouse;  2) Multidimensional  OLAP
(MOLAP),  which uses  array-based  multidimensional storage  engines for  multidimensional
views of data, rather than in a  relational database.  It often requires pre-processing to
create  data cubes.  3) Hybrid  OLAP (HOLAP),  which is  a combination  of both  ROLAP and
MOLAP.  It uses  both relational  and multidimensional  techniques to  inherit the  higher
scalability of ROLAP and the faster computation of MOLAP.

In the context of agricultural Big Data, HOLAP  is more suitable than both ROLAP and MOLAP
because: 1)  ROLAP has  quite slow  performance and does  not meet  all the  users' needs,
especially  when performing  complex calculations;  2) MOLAP  is not  capable of  handling
detailed  data  and  requires all  calculations  to  be  performed  during the  data  cube
construction; 3) HOLAP inherits  advantages of both ROLAP and MOLAP,  which allow the user
to  store large  data volumes  of detailed  information and  perform complex  calculations
within reasonable response time. 

\section{Quality Criteria}
\label{sec:QC}

The accuracy of data  mining and analysis techniques depends on the quality  of the DW. As
mentioned in \cite{Adelman-Moss.2000} and  \cite{Kimball-Ross.2013}, to build an efficient
ADW, the quality of the DW should meet the following important criteria:
\begin{enumerate}
  \item Making information easily accessible.
  \item Presenting consistent information.
  \item Integrating data correctly and completely.
  \item Adapting to change.
  \item Presenting and providing right information at the right time.
  \item Being a secure bastion that protects the information assets.
  \item Serving as  the authoritative  and trustworthy  foundation for  improved decision
        making. The analytics tools need to provide right information at the right time.
  \item Achieving benefits, both tangible and intangible.
  \item Being accepted by DW users.
\end{enumerate}

The above criteria must be formulated in a form of measurements. For example, with the 8th
criterion,  it needs  to determine  quality indicators  about benefits,  such as  improved
fertiliser management,  cost containment, risk  reduction, better or faster  decision, and
efficient  information transaction.  In the  last  criterion, a  user satisfaction  survey
should be used to find out how a given DW satisfies its user’s expectations.
 
\section{ADW Implementation}
\label{sec:ADWI}
Currently,  there  are  many  popular   large-scale  database  types  that  can  implement
DWs. Redshift \citep{Amazon.2018},  Mesa \citep{Gupta.2016}, Cassandra \citep{Hewitt.2016,
  Neeraj.2015},   MongoDB  \citep{Chodorow.2013,   Hows.2015}  and   Hive  \citep{Du.2018,
  Lam.2016}.  In \cite{Ngo.2019}, the authors analysed the most popular no-sql databases,
which fulfil most  of the aforementioned criteria. The advantages,  disadvantages, as well
as  similarities and  differences between  Cassandra, MongoDB  and Hive  were investigated
carefully in  the context  of ADW.  It was  reported that Hive  is a  better choice  as it
can be paired with MongoDB to implement the proposed ADW for the following reasons:

\begin{enumerate}
  \vspace{-1mm}
  \item Hive is  based on Hadoop which  is the most powerful cloud  computing platform for
        Big Data.   Besides, HQL is similar  to SQL which  is popular for the  majority of
        users. Hive  supports well  high storage capacity,  business intelligent  and data
        science  more  than MongoDB  or  Cassandra.  These  Hive  features are  useful  to
        implement ADW.
  \item Hive does not  have real-time performance so it needs to  be combined with MongoDB
        or Cassandra to improve its performance.
  \item MongoDB  is more suitable  than Cassandra to  complement Hive because:  1) MongoDB
        supports joint  operation, full text search,  ad-hoc query and second  index which
        are helpful to interact with the users. Cassandra does not support these features;
        2)  MongoDB has  the same  master –  slave  structure with  Hive that  is easy  to
        combine. While the structure of Cassandra is peer - to - peer; 3) Hive and MongoDB
        are more  reliable and  consistent. So  the combination of  both Hive  and MongoDB
        adheres to the CAP theorem.
\end{enumerate}

\begin{figure*}
  \vspace{-4mm}
  \begin{center}
    \includegraphics[width=0.76\textwidth]{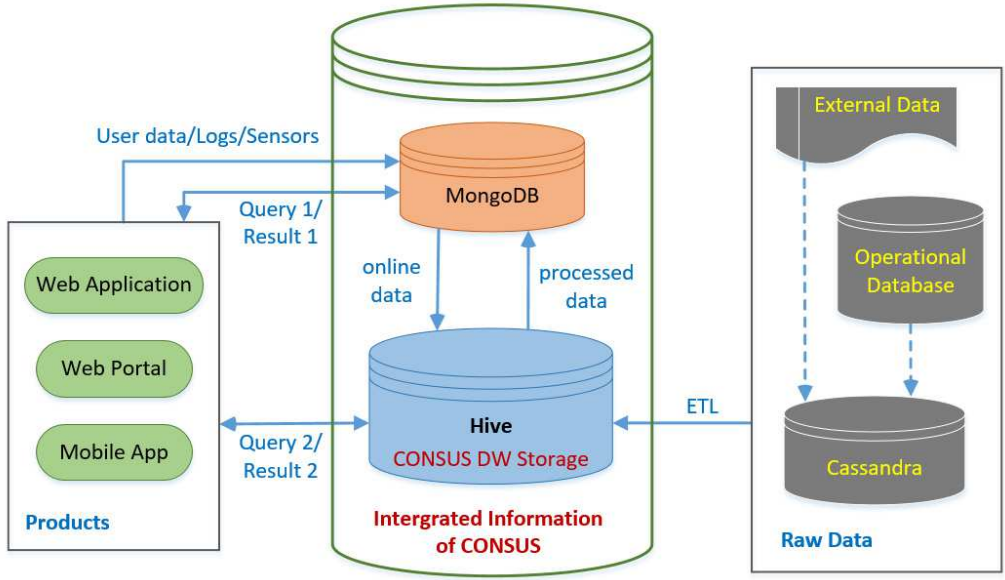}
  \end{center}
  \vspace{-4mm}
  \caption{Agricultural Data Warehouse Implementation}
  \label{fig_DW_Storage}
  \vspace{-2mm}
\end{figure*}	
	
The ADW implementation is illustrated  in Figure \ref{fig_DW_Storage} which contains three
modules, namely Integrated Information, Products  and Raw Data. The Integrated Information
module includes two components; MongoDB and Hive. MongoDB receives real-time data; as user
data, logs,  sensor data  or queries from  Products module, such  as web  application, web
portal or mobile app. Besides, some results which need to be obtained in real-time will be
transferred  from the  MongoDB to  Products. Hive  stores the  online data  and sends  the
processed data to MongoDB.  Some kinds of queries having complex calculations will be sent
directly to Hive. 

In the Raw Data module, almost data  in Operational Databases or External Data components,
is  loaded  into  Cassandra.  It  means  that we  use  Cassandra  to  represent  raw  data
storage. Hence, with the  diverse formats of raw data; image,  video, natural language and
sql data, Cassandra is  better to store them than SQL databases. In  the idle times of the
system, the  updated raw  data in  Cassandra will be  imported into  Hive through  the ELT
tool. This improves the performance of ETL and helps us deploy ADW on cloud or distributed
systems.  

\section{Performance Analysis}
\label{sec:PA}

The performance  analysis was conducted using  MySQL 5.7.22, JDK 1.8.0\_171,  Hadoop 2.6.5
and Hive 2.3.3  which run on Bash, on  Ubuntu 16.04.2, and on Windows  10. All experiments
were run  on a desktop with  an Intel Core  i7 CPU (2.40 GHz)  and 16 GB memory.   We only
evaluate  the performance  of reading  operation as  ADW is  used for  reporting and  data
analysis.  The database of ADW is duplicated into MySQL to compare performance. By
combining popular  HQL/SQL commands, namely  Where, Group  by, Having, Left  (right) Join,
Union and Order by,  we created 10 groups for testing. Every group  has 5 queries and uses
one,  two or  more commands  (see Table  \ref{tab_queries}).  Moreover,  every query  uses
operators; And, Or, $\ge$, Like, Max, Sum and Count, to express complex queries.
		
\begin{table}[H]
  \caption{Command combinations of queries}
  \small
  \begin{tabular}{c|l}
    \textbf{Group} & \textbf {Commands} \\
    \hline
    $G_1$ & Where\\
    $G_2$ & Where, Group by \\
    $G_3$ & Where, Left (right) Join \\
    $G_4$ & Where, Union \\
    $G_5$ & Where, Order by \\
    $G_6$ & Where, Left (right) Join, Order by \\
    $G_7$ & Where, Group by, Having \\
    $G_8$ & Where, Group by, Having, Order by \\
    $G_9$ & Where, Group by, Having, Left (right) Join, \\ 
          & Order by \\
    $G_{10}$ & Where, Group by, Having, Union, Order by \\
  \end{tabular}
  \vspace{-2mm}
  \label{tab_queries}
\end{table}

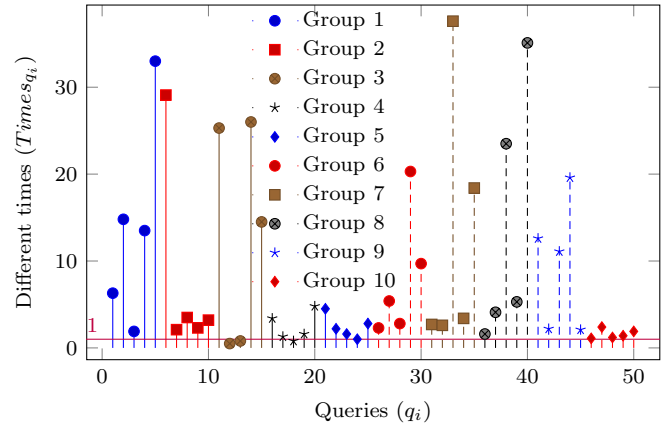
\begin{figure}[H]
  \begin{center}
    \begin{tikzpicture}
      \pgfplotsset{height=63mm, width= 92mm, xlabel near ticks, ylabel near ticks}
      \begin{axis}[
	font=\footnotesize,
	enlargelimits=0.05,
	legend style={draw=none, fill=none, font=\footnotesize, cells={anchor=west} , legend pos=north east, xshift= -31mm,yshift= 2mm }, xlabel={Queries ($q_i$)}, ylabel={Different times ($Times_{q_i}$)},
	]
	\addplot+[ycomb] plot coordinates {(1, 6.3) (2, 14.8) (3, 1.9) (4, 13.5) (5, 33) };
	\addplot+[ycomb] plot coordinates {(6, 29.1) (7, 2.1) (8, 3.5) (9, 2.3) (10, 3.2) };
	\addplot+[ycomb] plot coordinates {(11, 25.3) (12, 0.5) (13, 0.8) (14, 26) (15, 14.5) };
	\addplot+[ycomb] plot coordinates {(16, 3.4) (17, 1.3) (18, 0.8) (19, 1.6) (20, 4.8) };
	\addplot+[ycomb] plot coordinates {(21, 4.5) (22, 2.2) (23, 1.6) (24, 1) (25, 2.8) };
	\addplot+[ycomb] plot coordinates {(26, 2.3) (27, 5.4) (28, 2.8) (29, 20.3) (30, 9.7) };
	\addplot+[ycomb] plot coordinates {(31, 2.7) (32, 2.6) (33, 37.6) (34, 3.4) (35, 18.4) };
	\addplot+[ycomb] plot coordinates {(36, 1.6) (37, 4.1) (38, 23.5) (39, 5.3) (40, 35.1) };
	\addplot+[ycomb] plot coordinates {(41, 12.6) (42, 2.2) (43, 11.1) (44, 19.6) (45, 2.1) };
	\addplot+[ycomb] plot coordinates {(46, 1.1) (47, 2.4) (48, 1.2) (49, 1.4) (50, 1.9) };
	
	\addplot[purple,sharp plot,update limits=false]
	coordinates {(-2,1) (53,1)}
	node[above] at (axis cs:-0.9,0.8) {1};
	
	\legend{Group 1, Group 2, Group 3, Group 4, Group 5, Group 6, Group 7, Group 8, Group 9, Group 10}
      \end{axis}
    \end{tikzpicture}
  \end{center}
  \captionsetup{justification=centering}
  \vspace{-4mm}
  \caption{Different times between MySQL and \\ ADW in runtime of every Query}
  
  \label{fig:times50queries}
  \vspace{-2mm}
\end{figure}

All queries were executed three times and we took the average value of the their execution
timess. The difference in runtime between MySQL and ADW for a query $q_i$ is calculated as
$Times_{q_i}   =    RT^{mysql}_{q_i}/RT^{ADW}_{q_i}$.   Where,    $RT^{mysql}_{q_i}$   and
$RT^{ADW}_{q_i}$   are   average   runtimes   of   query   $q_i$   on   MySQL   and   ADW,
respectively. Moreover, with each group $G_i$, the difference in runtime between MySQL and
ADW is $Times_{G_i} = RT^{mysql}_{G_i}/RT^{ADW}_{G_i}$. Where, $RT_{G_i} = Average(RT_{q_i})$
is average runtime of group $G_i$ on MySQL or ADW.

Figure \ref{fig:times50queries}  describes the time  difference between MySQL and  ADW for
every query. Although running  on one computer, but with large data  volume, ADW is faster
than MySQL on 46 out of 50 queries. MySQL is faster for three queries $12^{th}$, $13^{th}$
and $18^{th}$ belonging to groups $3^{rd}$ and $4^{th}$. The two systems returned the same
time  for  query  $24^{th}$  from  group  $5^{th}$. Within  each  query  group,  for  fair
performance comparison, the  queries combine randomly fact tables  and dimensional tables.
This makes complex queries  taking more time and the time  difference is significant. When
varying the sizes  and structures of the  tables, the difference is  very significant; see
Figure \ref{fig:times50queries}.
	
		
  \begin{figure}[H]
    \vspace{0mm}
    \begin{center}
    \begin{tikzpicture}
    \pgfplotsset{height=52mm, width= 72mm, xlabel near ticks, ylabel near ticks}
    \begin{axis}[
    font=\footnotesize,
    enlargelimits=0.2,
    legend style={draw=none, fill=none, font=\footnotesize, cells={anchor=west} , legend pos=north east, xshift= 25mm,yshift= 2mm }, 
    xlabel={Groups ($G_i$)}, ylabel={Different times ($Times_{G_i}$)},
    nodes near coords,
    nodes near coords align={vertical},
    ]
    
    \addplot+[ycomb] plot coordinates {(1, 6.24)};
    \addplot+[ycomb] plot coordinates {(2, 2.92)};
    \addplot+[ycomb] plot coordinates {(3, 1.22)};
    \addplot+[ycomb] plot coordinates {(4, 2.86)};
    \addplot+[ycomb] plot coordinates {(5, 2.27)};
    \addplot+[ycomb] plot coordinates {(6, 4.66)};
    \addplot+[ycomb] plot coordinates {(7, 3.36)};
    \addplot+[ycomb] plot coordinates {(8, 4.63)};
    \addplot+[ycomb] plot coordinates {(9, 3.16)};
    \addplot+[ycomb] plot coordinates {(10, 1.56)};
    
    \addplot[purple,sharp plot,update limits=false]
    coordinates {(-0.1,3.19) (12,3.19)}
    node[above] at (axis cs:0, 2.5) {Mean};
    
    \end{axis}
    \end{tikzpicture}
    \end{center}
    
    \vspace{-5mm}
    \centering
    \captionsetup{justification=centering}
    \caption{
    Different times between MySQL and \\ADW in runtime of every group}
    
    \label{fig:times10groups}
    \vspace{-3mm}
\end{figure}
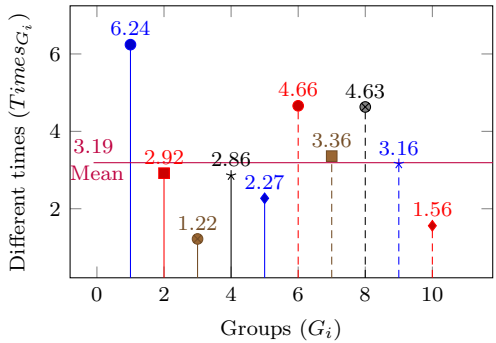

Beside comparing runtime in every query, we  aslo compare runtime of every group presented
in Figure  \ref{fig:times10groups}. Comparing  to MySQL,  ADW is more  than at  most (6.24
times) at group $1^{st}$ which uses only \textit{Where} command, and at least (1.22 times)
at group $3^{rd}$ which uses \textit{Where} and \textit{Joint} commands. 

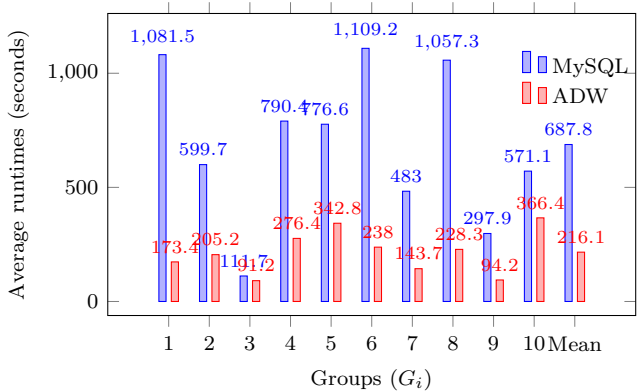
\begin{figure}[H]
	\vspace{-2mm}
	\begin{center}
	\begin{tikzpicture}
	\pgfplotsset{height=56mm, width= 86mm, xlabel near ticks, ylabel near ticks}
	\begin{axis}[
	    font=\footnotesize,
	    ybar,
	    bar width=2.7pt,
	    enlargelimits=0.15,
	    legend style={draw=none, fill=none, font=\footnotesize, cells={anchor=west} ,  legend pos=north east,  xshift= 3mm,yshift= -3mm}, 
	    xlabel={Groups ($G_i$)},
	    ylabel={Average runtimes (seconds)},
	    symbolic x coords={1, 2, 3, 4, 5, 6, 7, 8, 9, 10, Mean },
	    xtick=data,
	    nodes near coords,
		 every node near coord/.append style={font=\scriptsize},		    
	    ]
		
	\addplot coordinates {(1, 1081.5) (2, 599.7) (3, 111.7) (4, 790.4) (5, 776.6) (6, 1109.2) (7, 483) (8, 1057.3) (9, 297.9) (10, 571.1) (Mean, 687.8)};
	\addplot coordinates {(1, 173.4) (2, 205.2) (3, 91.2) (4, 276.4) (5, 342.8) (6, 238) (7, 143.7) (8, 228.3) (9, 94.2) (10, 366.4) (Mean, 216.1)};
			
	\legend{MySQL, ADW}
	\end{axis}
	\end{tikzpicture}
	\end{center}
	\vspace{-5mm}
	\captionsetup{justification=centering}
	\caption{Average Runtimes of MySQL and \\ ADW in every Groups}
	\label{fig:runtime_groups}
	\vspace{-2mm}
\end{figure}

Figure \ref{fig:runtime_groups}  presents the average  runtime of  the 10 query  groups on
MySQL and ADW. Mean, the run time of a reading query on MySQL and ADW is 687.8 seconds and
216.1 seconds, respectively.   It means that ADW  is faster 3.19 times. In  the future, by
deploying ADW  solution on cloud or  distributed systems, we believe  that the performance
will be even much better than MySQL.

\section{Application for Decision Making}
\label{sec:ADM}

The proposed ADW and study its performance  on real agricultural data, we illustrated some
queries examples to  show how to extract information from  ADW.  These queries incorporate
inputs  on  crop, yield,  pest,  soil,  fertiliser,  inspection, farmer,  businessman  and
operation time to reduce labour and  fertiliser inputs, farmer services, disease treatment
and also increase yields.  These query information  could not be extracted if the Origin's
separate 29 datasets have  not been integrated into ADW. The  data integration through ADW
is  actually  improve  the   value  of  a  crop  management  data   over  time  to  better
decision-making.

\textbf{Example 1}: List fields, crops in the fields, yield and pest in the field with conditions: (1) the fields do not used 'urea' fertilizer; (2) the crops has 'yellow rust' or 'brown rust' diseases; (3) the crops were grown in 2015.
\small
\begin{verbatim}
select CR.CropName, FI.FieldName, FF.Yield,
       PE.CommonName, FF.PestNumber, PE.Description
from FieldFact FF, Crop CR, Field FI, Pest PE,
     Fertiliser FE, Inspection INS, OperationTime OP
where FF.CropID = CR.CropID and
      FF.FieldID = FI.FieldID and
      FF.PestID = PE.PestID and
      FF.FertiliserID = FE.FertiliserID and
      CR.CropID = INS.CropID and
      FF.OperationTimeID = OP.OperationTimeID and
      FE.FertiliserName <> 'urea' and
      (INS.Description = 'Yellow Rust' or 
             INS.Description = 'Brown Rust') and
      Year(INS.Date) = '2015' and
      Year(OP.StartDate) = '2015' and 
      Year(OP.EndDate) = '2015'
\end{verbatim}
\normalsize

\textbf{Example 2}: List farmers and their crop quantities were sold by Ori Agro company in 08/2016. 
\small
\begin{verbatim}
select FA.FarmerID, FA.FarmerName, CR.CropName, 
       SF.Unit, SUM(SF.Quantity)
from Salefact SF, business BU, farmer FA, crop CR
where SF.BusinessID = BU.BusinessID and 
      SF.FarmerID = FA.FarmerID and
      SF.CropID = CR.CropID and
      Month(SF.SaleDate) = '08' and
      Year(SF.SaleDate) = '2016' and
      BU.BusinessName = 'Ori Agro'
group by CR.CropName
\end{verbatim}
\normalsize

\textbf{Example 3}: List Crops and their fertiliser and treatment information. In that, crops were cultivated and harvested in 2017, Yield $>$ 10 tons/ha and attached by 'black twitch' pest. Besides, the soil in field has PH $> 6$ and Silt $<=50$ mg/l.

\small
\begin{verbatim}
Select CR.CropName, FE.FertiliserName,
       FF.FertiliserQuantity, TR.TreatmentName,
       TR.Rate, TR.TreatmentComment
From FieldFact FF, Crop CR, OperationTime OT,
     Soil SO, PEST PE, Fertiliser FE, Treatment TR
Where FF.CropID = CR.CropID and
      FF.OperationTimeID = OT.OperationTimeID and
      FF.SoildID = SO.SoilID and
      FF.PestID = PE.PestID and
      FF.FertiliserID = FE.FertiliserID and
      FF.TreatmentID = TR.TreatmentID and
      Year(OT.StartDate) = '2017' and
      Year(OT.EndDate) = '2017' and
      FF.Yield > 10 and
      SO.PH > 6 and SO.Silt <= 50 and
      PE.CommonName = 'Black twitch'
\end{verbatim}
\normalsize

\textbf{Example 4}: List crops, fertilisers, corresponding fertiliser quantities in spring, 2017 in every field and site of 10 farmers (crop companies) who used the large amount of $P_2O_5$ in winter, 2016.
\\

To execute this request, the query needs to exploit data in the FieldFact fact table and the six dimension tables, namely Crop, Field, Site, Farmer, Fertiliser and OperationTime. The query consists of two subqueries which return \textit{10 farmers (crop companies) that used the largest amount of Urea in spring, 2016}.

\small
\begin{verbatim}
Select FI.FieldName, SI.SiteName, FA.FarmerName, 
       CR.CropName, FE.FertiliserName, 
       FF.FertiliserQuantity, FE.Unit, OT.StartDate
From FieldFact FF, Crop CR, Field FI, Site SI, 
     Farmer FA, Fertiliser FE, Operationtime OT
Where FF.CropID = CR.CropID and 
      FF.FieldID = FI.FieldID and 
      FF.FertiliserID = FE.FertiliserID and
      FF.OperationTimeID = OT.OperationTimeID and 
      FI.SiteID = SI.SiteID and 
      SI.FarmerID = FA.FarmerID and 
      OT.Season = 'Spring' and 
      YEAR(OT.StartDate) = '2017' and 
      FA.FarmerID IN(
       Select FarmerID
       From
       (Select SI.FarmerID as FarmerID, 
         SUM(FF.FertiliserQuantity) as SumFertiliser
        From FieldFact FF, Field FI, Site SI,
             Fertiliser FE, OperationTime OT
        Where FF.FieldID = FI.FieldID and 
              FF.FertiliserID = FE.FertiliserID and
              FF.OperationTimeID = 
                             OT.OperationTimeID and
              SI.SiteID = FI.SiteID and
              FE.FertiliserName = 'SO3' and
              OT.Season = 'Spring' and 
              YEAR(OT.StartDate) = '2016'
         Group by SI.FarmerID
         Order by SumFertiliser DESC
         Limit 10
        )AS Table1
      )
\end{verbatim}
\normalsize

\section{Conclusion and Future Work}
\label{sec:CFW}

In this paper, we presented a schema herein optimised for the real agricultural datasets that were made available to us. The schema been designed as a constellation so it is flexible to adapt to other agricultural datasets and quality criteria of agricultural Big Data. Based on some existing popular open source DWs,
We designed and implemented the agricultural DW by combining Hive, MongoDB and Cassandra DWs to exploit their advantages and overcome their limitations. ADW includes necessary modules to deal with large scale and efficient analytics for agricultural Big Data. Moreover, through particular reading queries using popular HQL/SQL commands,  ADW storage outperforms MySQL by far. Finally, we outlined some complex HQL queries that enabled knowledge extraction from ADW to optimize of agricultural operations.

In the future work, we shall pursue the deployment of ADW on a cloud system and implement more functionalities to exploit this DW. The future developments will include: (1) experimentation and analyzation the performance of MongoDB and the affectation between MongoDB and Hive; (2) The sophisticated the data mining and the spreading activation algorithms \citep{Ngo.2014} to determine crop data characteristics and combine with expected outputs to extract useful knowledge; (3) Predictive models based on machine learning algorithms; (4) An intelligent interface and graph representation \citep{Helmer.2015} for data access; (5) Combination with the ontology to extract knowledge \citep{Ngo.2011, Cao.2012}.

\section*{Appendix}
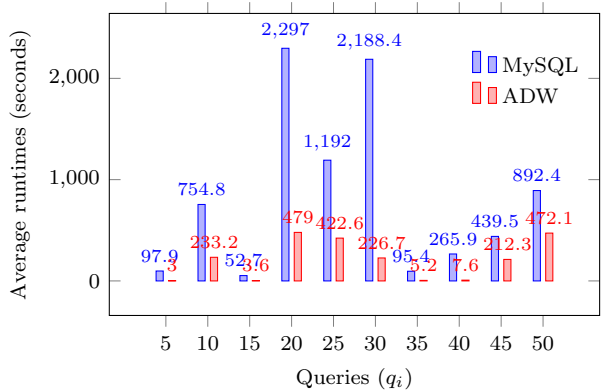
\begin{figure}[H]
	\begin{center}
	\begin{tikzpicture}
	\pgfplotsset{height=56mm, width= 81mm, xlabel near ticks, ylabel near ticks}
	\begin{axis}[
	    font=\footnotesize,
	    ybar,
	    bar width=2.7pt,
	    enlargelimits=0.15,
	    legend style={draw=none, fill=none, font=\footnotesize, cells={anchor=west} ,  legend pos=north east,  xshift= 1mm,yshift= -3mm}, 
	    xlabel={Queries ($q_i$)},
	    ylabel={Average runtimes (seconds)},
	    symbolic x coords={5, 10, 15, 20, 25, 30, 35, 40, 45, 50},
	    xtick=data,
	    nodes near coords,
		 every node near coord/.append style={font=\scriptsize},		    
	    ]
		
	\addplot coordinates {(5, 97.9) (10, 754.8) (15, 52.7) (20, 2297) (25, 1192) (30, 2188.4) (35, 95.4) (40, 265.9) (45, 439.5) (50, 892.4)};
	\addplot coordinates {(5, 3) (10, 233.2) (15, 3.6) (20, 479) (25, 422.6) (30, 226.7) (35, 5.2) (40, 7.6) (45, 212.3) (50, 472.1)};
			
	\legend{MySQL, ADW}
	\end{axis}
	\end{tikzpicture}
	\end{center}
	\vspace{-5mm}
	\captionsetup{justification=centering}
	\caption{Average runtimes of MySQL and \\ ADW in 10 typical queries}
	\label{fig:runtime_10queries_belong10groups}
	\vspace{-2mm}
\end{figure}

The followings are HQL/SQL scripts of 10 queries which are representative of 10 query groups. The average runtimes of these queries on MySQL and ADW are shown in Figure ~\ref{fig:runtime_10queries_belong10groups}.

\noindent
1) The query $5^{th}$ belongs to the group $1^{st}$:
\vspace{-2mm}
\small
\begin{verbatim}
SELECT fieldfact.FieldID, crop.cropname,
       fieldfact.yield
FROM fieldfact, crop
WHERE fieldfact.cropid = crop.cropid and 
      SprayQuantity = 7 and 
      (crop.CropName like 'P\%' or 
       crop.CropName like 'R\%' or 
       crop.CropName like 'G\%');
\end{verbatim}

\noindent \normalsize
2) The query $10^{th}$ belongs to the group $2^{nd}$:
\vspace{-2mm}
\small
\begin{verbatim}
SELECT soil.PH, count(*)
FROM fieldfact, soil
WHERE fieldfact.SoildID = soil.SoilID and 
      fieldfact.sprayquantity = 2  
GROUP by soil.PH; 
\end{verbatim}

\noindent \normalsize
3) The query $15^{th}$ belongs to the group $3^{rd}$:
\vspace{-2mm}
\small
\begin{verbatim}
SELECT fieldfact.yield,
       fertiliser.fertiliserName,
       fertiliser.fertiliserGroupName
FROM fieldfact
RIGHT JOIN fertiliser on 
     fieldfact.fertiliserID = fertiliser.fertiliserID 
WHERE fieldfact.fertiliserQuantity = 10 and
      fertiliser.fertiliserName like '%slurry%';
\end{verbatim}

\noindent \normalsize
4) The query $20^{th}$ belongs to the group $4^{th}$:
\vspace{-2mm}
\small
\begin{verbatim}
SELECT sprayproductname
FROM fieldfact, spray
WHERE fieldfact.sprayid = spray.sprayid and
      fieldfact.watervolumn > 5 and
      fieldfact.watervolumn < 20
UNION
SELECT productname
FROM product, orderfact
WHERE product.ProductID = orderfact.ProductID 
      and (orderfact.Quantity = 5 or 
           orderfact.Quantity = 6);
\end{verbatim}

\noindent \normalsize
5) The query $25^{th}$ belongs to the group $5^{th}$:
\vspace{-2mm}
\small
\begin{verbatim}
SELECT fieldfact.fieldID, field.FieldName, 
       field.FieldGPS, spray.SprayProductName
FROM fieldfact, field, spray
WHERE fieldfact.FieldID = field.FieldID and
      fieldfact.SprayID = spray.SprayID and
      fieldfact.PestNumber = 6
ORDER BY field.FieldName;
\end{verbatim}

\noindent \normalsize
6) The query $30^{th}$ belongs to the group $6^{th}$:
\vspace{-2mm}
\small
\begin{verbatim}
SELECT fieldfact.FieldID, nutrient.NutrientName,
       nutrient.Quantity, nutrient.`Year`
FROM fieldfact
RIGHT JOIN nutrient on 
     fieldfact.NutrientID = nutrient.NutrientID
WHERE fieldfact.NutrientQuantity = 3 and
      fieldfact.fertiliserquantity = 3
ORDER BY nutrient.NutrientName
LIMIT 10000;
\end{verbatim}

\noindent \normalsize
7) The query $35^{th}$ belongs to the group $7^{th}$:
\vspace{-2mm}
\small
\begin{verbatim}
SELECT crop.cropname,
       sum(fieldfact.watervolumn) as sum1
FROM fieldfact, crop
WHERE fieldfact.cropid = crop.cropid and 
      fieldfact.sprayquantity = 8 and 
      crop.EstYield >= 1 and crop.EstYield <=10
GROUP BY crop.cropname
HAVING sum1 > 100;
\end{verbatim}

\noindent \normalsize
8) The query $40^{th}$ belongs to the group $8^{th}$:
\vspace{-2mm}
\small
\begin{verbatim}
SELECT crop.cropname, 
       sum(fieldfact.fertiliserquantity) as sum1
FROM fieldfact, crop
WHERE fieldfact.cropid = crop.cropid and
      fieldfact.nutrientquantity= 5 and
      crop.EstYield <=1
GROUP by crop.cropname
HAVING sum1 > 30
ORDER BY crop.cropname;
\end{verbatim}

\noindent \normalsize
9) The query $45^{th}$ belongs to the group $9^{th}$:
\vspace{-2mm}
\small
\begin{verbatim}
SELECT nutrient.NutrientName,
       sum(nutrient.Quantity) as sum1
FROM fieldfact
LEFT JOIN nutrient on 
    fieldfact.NutrientID = nutrient.NutrientID
WHERE nutrient.nutrientName like '%tr%' and
     (fieldfact.pestnumber = 16 or
      fieldfact.pestnumber = 15)
GROUP by nutrient.NutrientName
HAVING sum1 <300
ORDER BY nutrient.NutrientName;
\end{verbatim}

\noindent \normalsize
10) The query $50^{th}$ belongs to the group $10^{th}$:
\vspace{-2mm}
\small
\begin{verbatim}
SELECT sprayproductname as name1,
       sum(fieldfact.watervolumn) as sum1
FROM fieldfact, spray
WHERE fieldfact.sprayid = spray.sprayid and
      fieldfact.Yield > 4 and fieldfact.Yield < 8
GROUP by sprayproductname
HAVING sum1 > 210
UNION
SELECT productname as name1,
       sum(orderfact.Quantity) as sum2
FROM product, orderfact
WHERE product.ProductID = orderfact.ProductID and
     (orderfact.Quantity = 5 or 
      orderfact.Quantity = 6)
GROUP by productname
HAVING sum2 > 50
ORDER BY name1;
\end{verbatim}

\vspace{-2mm}
\section*{Acknowledgment}
\vspace{-3mm}

This research is an extended work of \cite{Ngo.2019} being part of the CONSUS research program. It is funded under the SFI Strategic Partnerships Programme (16/SPP/3296) and is co-funded by Origin Enterprises Plc.

\bibliography{mybibfile}
\bibliographystyle{apa.bst}

\end{document}